\documentclass[%
 reprint,
 amsmath,amssymb,
 aps,
prl
]{revtex4-1}

\usepackage{dcolumn}
\usepackage{bm}

\usepackage{amsmath, amssymb}
\usepackage{amsfonts}
\usepackage{braket}

\usepackage{ascmac}

\usepackage{graphicx}

\newcommand{\eq}[1]{\begin{equation} #1 \end{equation}}
\newcommand{\eqa}[2]{\begin{equation} #1 \label{#2} \end{equation}}
\newcommand{\balign}[1]{\begin{align} #1 \end{align}}

\newcommand{\figin}[4]
{\begin{figure}[tb]
\centering
\includegraphics[width= #1]{#2.pdf}
\caption{#3}
\label{f:#4}
\end{figure}}

\newcommand{\todayd}{\the\year/\the\month/\the\day}

\newcommand{\bib}{\bibitem}

\newcommand{\up}{\uparrow}
\newcommand{\down}{\downarrow}

\newcommand{\Lmd}{\Lambda}

\newcommand{\lb}{\label}
\newcommand{\nt}{\notag}
\newcommand{\Tr}{\mathrm{Tr}}

\newcommand{\bel}{\begin{easylist}}
\newcommand{\eel}{\end{easylist}}

\newcommand{\eref}[1]{Eq.~\eqref{#1}}

\newcommand{\fref}[1]{Fig.~\ref{f:#1}}

\def \({\left(}
\def \){\right)}
\newcommand{\la}{\left\langle}
\newcommand{\ra}{\right\rangle}
\newcommand{\abs}[1]{\left|#1\right|}

\newcommand{\sumtwo}[2]%
{\mathop{\sum_{#1}}_{#2}}
\newcommand{\sumthree}[3]%
{\mathop{\mathop{\sum_{#1}}_{#2}}_{#3}}
\newcommand{\sumfour}[4]%
{\mathop{\mathop{\mathop{\sum_{#1}}_{#2}}_{#3}}_{#4}} 
\newcommand{\prodtwo}[2]%
{\mathop{\prod_{#1}}_{#2}}
\newcommand{\mintwo}[2]%
{\mathop{\min_{#1}}_{#2}}
\newcommand{\maxtwo}[2]%
{\mathop{\max_{#1}}_{#2}}
\newcommand{\maxthree}[3]%
{\mathop{\mathop{\max_{#1}}_{#2}}_{#3}}
\newcommand{\limtwo}[2]%
{\mathop{\lim_{#1}}_{#2}}
\newcommand{\suptwo}[2]%
{\mathop{\sup_{#1}}_{#2}}
\newcommand{\supthree}[3]%
{\mathop{\mathop{\sup_{#1}}_{#2}}_{#3}}
\newcommand{\supfour}[4]%
{\mathop{\mathop{\mathop{\sup_{#1}}_{#2}}_{#3}}_{#4}} 
\newcommand{\inftwo}[2]%
{\mathop{\inf_{#1}}_{#2}}
\newcommand{\infthree}[3]%
{\mathop{\mathop{\inf_{#1}}_{#2}}_{#3}}
\newcommand{\inffour}[4]%
{\mathop{\mathop{\mathop{\inf_{#1}}_{#2}}_{#3}}_{#4}} 

\newcommand\calH{{\cal H}}

\newcommand\calQ{{\cal Q}}

\newcommand\calT{{\cal T}}

\newcommand{\bsS}{\boldsymbol{S}}

\newcommand{\ep}{\varepsilon}

\newcommand{\Di}{\mathit{\Delta}}

\newcommand{\para}[1]{{\em #1}\/.---}

\def\rnum#1{\resizebox{0.5em}{\height}{\expandafter{\romannumeral #1}}}
\def\Rnum#1{\resizebox{0.5em}{\height}{\uppercase\expandafter{\romannumeral #1}}}
\newcommand{\hP}{\hat{P}}
\newcommand{\hh}{\hat{h}}
\newcommand{\hH}{\hat{H}}
\newcommand{\hS}{\hat{S}}
\newcommand{\hbS}{\hat{\bsS}}

\newcommand{\Sa}{\hat{S}^\alpha}
\newcommand{\ts}{\tilde{\sigma}}
\newcommand{\hcQ}{\hat{\calQ}}
\newcommand{\hQ}{\hat{Q}}
\newcommand{\hO}{\hat{O}}

\begin{document}

\preprint{APS/123-QED}

\title{Systematic Construction of Counterexamples to Eigenstate Thermalization Hypothesis}

\author{Naoto Shiraishi}
\affiliation{%
Department of Physics, Keio University, 3-14-1 Hiyoshi, Yokohama 223-8522, Japan
}%

\author{Takashi Mori}%
\affiliation{Department of Physics, The University of Tokyo,7-3-1 Hongo, Bunkyo-ku, Tokyo 113-0033, Japan}

\date{\today}

\begin{abstract}
We propose a general method to embed target states into the middle of the energy spectrum of a many-body Hamiltonian as its energy eigenstates.
Employing this method, we construct a translationally-invariant local Hamiltonian with no local conserved quantities, which does not satisfy the eigenstate thermalization hypothesis.
The absence of eigenstate thermalization for target states is analytically proved and numerically demonstrated.
In addition, numerical calculations of two concrete models also show that all the energy eigenstates except for the target states have the property of eigenstate thermalization, from which we argue that our models thermalize after a quench even though they does not satisfy the eigenstate thermalization hypothesis.
\begin{description}
\item[PACS numbers]
05.30.-d, 
05.70.Ln, 
03.65.-w, 
75.10.Pq	
\end{description}
\end{abstract}

\pacs{Valid PACS appear here}

\maketitle

\para{Introduction}
The emergence of the arrow of time, in particular thermalization to the equilibrium state, in macroscopic many-body systems from reversible microscopic dynamics is one of the most important and profound problems in theoretical physics.
This problem has been discussed since the early days of statistical mechanics.
One important observation is that almost all pure states in the energy shell of any given energy are {\it thermal}.
Here, a quantum state is said to be thermal when the expectation value of any local observable in this state coincides with that obtained by the corresponding microcanonical ensemble within a certain small error vanishing in the thermodynamic limit.
Fragments of this idea have already seen in Boltzmann~\cite{Bol}, von Neumann~\cite{Neu}, Tolman~\cite{Tol}, Khinchin~\cite{Khi}, and Schr\"{o}dinger~\cite{Sch},  and now this idea is known as {\it typicality} of thermal equilibrium~\cite{BL, Llo, PSW, Gold, Sug, Rei07}.
Although typicality is widely believed to provide a satisfactory characterization of thermal equilibrium, it is far from sufficient to explain thermalization~\cite{Tas16}.
 (The precise definition of thermalization is given in \cite{SM}.)

It has been well established that thermalization in an isolated quantum system can be explained under the assumption that {\it every} energy eigenstate is thermal~\cite{note1}.
This assumption is referred to as the {\it eigenstate thermalization hypothesis} (ETH)~\cite{Neu, Deu, Sre, HZB, Tas98, Rig08}.
(The precise definition of the ETH is given in \cite{SM}.)

It is remarked that the property known as {\it weak}-ETH~\cite{Bir}, which states that {\it almost all} energy eigenstates are thermal, is not enough to explain thermalization.
Indeed, the weak-ETH can be proved for a broad class of translationally-invariant systems regardless of their integrability~\cite{Bir, IKS, Mor16}, while observations in experiments~\cite{Gri, Lan} and numerical simulations~\cite{Cra08, Rig09, Bir} report that integrable systems~\cite{integrable} do not thermalize.
Absence of thermalization in an integrable system is attributed to an important weight of the initial state to atypical nonthermal energy eigenstates~\cite{Bir,Rig16,Tas-priv}.
Thus the weak-ETH does not guarantee thermalization, and therefore the ETH has been considered to be a key ingredient for thermalization~\cite{DAl, GE16, Pal}.

Many numerical simulations report that the ETH is valid if the Hamiltonian of a many-body quantum system satisfies the following three conditions: (i) translation invariance (in particular, no localization~\cite{localization}), (ii) no local conserved quantity, and (iii) local interactions~\cite{Rig08, Rig09, Bir, SR, KIH, SHP, BMH, Sor, Mon}.
Here, the word {\it local} stands for both few-body and short-range.
Interestingly, all known examples not satisfying the ETH violates at least one of (i)-(iii).
Integrable systems~\cite{Bir, Rig09, SHP, SR} and systems with local symmetries~\cite{Ham} violate (ii), and systems with Anderson localization~\cite{HSS, And} or many-body localization~\cite{BAA, Imb} violate (i).
It is noteworthy that these examples do not thermalize.
It may be then tempting to conjecture that the above (i)-(iii) are necessary and sufficient conditions for the validity of ETH and also for the system to thermalize.

In this Letter we construct counterexamples of this conjecture.
We first propose a general method of {\it embedding}, and then, by using this method, we construct two concrete models which satisfy the three conditions (i)-(iii), but can be proved rigorously to violate the ETH~\cite{SM}.
Moreover, we also argue that these models exhibit thermalization after a physically-plausible quench.
Our findings do not only reveal the richness of quantum many-body systems, but also lead to reconsideration of the conventional beliefs on the mechanism of thermalization.

\para{Method of embedding}
We here explain the procedure of embedding.
Consider a quantum system on a discrete lattice with a set of sites $\Lmd$ with Hilbert space $\calH$.
Let  $\hP_i$ ($i=1,2,\cdots ,N$) be arbitrary local projection operators on $\calH$ which do not necessarily commute with each other.
We usually take $N=O(|\Lmd|)$, in particular $N=|\Lmd|$.
We define a subspace $\calT\subset\calH$ as a subspace spanned by the set of states $\ket{\Psi}\in\calH$ satisfying
\eqa{
\hP_i\ket{\Psi}=0 
}{proj}
for any $i$.
We assume that $\calT$ contains at least one non-vanishing state.
The states in $\calT$ are target states to be embedded.

Let $\hh_i$ ($i=1,2,\dots,N$) be arbitrary local Hamiltonians $\hh_i$, and let $\hH'$ be a Hamiltonian which satisfies $[\hH',P_i]=0$ for $i=1,\dots,N$.
We then construct the desired Hamiltonian as
\eqa{
\hH:=\sum_i \hP_i \hh_i \hP_i +\hH'.
}{defH}
Since $\hP_i\hH \ket{\Psi}=\hP_i\hH' \ket{\Psi}=\hH'\hP_i \ket{\Psi}=0$ for $\ket{\Psi}\in \calT$, we find that $\calT$ is invariant under the map with $\hH$, and thus the Hamiltonian $\hH$ has $\dim \calT$ energy eigenstates within $\calT$.
For a special case that $\hH'=0$ and all the eigenvalues of $\hh_i$ are nonnegative, this Hamiltonian is regarded as a frustration-free Hamiltonian, which is seen in Ref.~\cite{Cha14}.
In general, the eigenenergies of the embedded states are in the middle of the energy spectrum, and this procedure can be regarded as a general method of embedding the target states $\calT$ into the middle of the energy spectrum of a nonintegrable local Hamiltonian.

An embedded state $\ket{\Psi}$ satisfying \eref{proj} is a highly anomalous state in the sense that a local projection operator $\hP_i$ takes exactly zero expectation value with no fluctuation, which is unexpected behavior in a thermal state.
This observation leads to a crucial result that the ETH is always violated regardless of $\{ \hh_i\}$.
In fact, in line with the above intuition, the violation of ETH is rigorously proven when $\hH'$ is also written as $\hH'=\sum_i\hh_i'$ with local terms $\{\hh_i'\}$, and both $\{\hh_i\}$ and $\{\hh_i'\}$ are bounded operators~\cite{SM}.

In the following, we express the eigenstates of $\hH$ as $\ket{\phi_j}$ with sorting them by energy ($E_{j-1}\leq E_j\leq E_{j+1}$).
We also write the number of total eigenstates and those in $\calT$ as $N_{\rm tot}$ and $N_{\rm ex}$, respectively.

\para{Model 1: two dimer states}
We now construct the first counterexample to the ETH.
Consider a one-dimensional spin chain of $S=1/2$ with even length $L$ with the periodic boundary condition.
The sites are labeled as $i=1,2\cdots ,L$, and we identify $i=0,-1$ to $i=L,L-1$, and $i=L+1, L+2$ to $i=1,2$.
The spin operator on the site $i$ is denoted by $\hbS_i$.
We introduce the total spin operator of sites $i-1$, $i$ and $i+1$ denoted by 
\eq{
\hbS_i^{\rm tot, 3}=\hbS_{i-1}+\hbS_i+\cdot \hbS_{i+1},
}
whose length $S_i^{\rm tot, 3}$ takes $3/2$ or $1/2$~\cite{length}.
We then set the projection operator $\hP_i$ as that into the subspace with $S_i^{\rm tot, 3}=3/2$, which we denote by $\hP_i^{S=3/2}$.
In terms of spin operators, $\hP_i^{S=3/2}$ is expressed as 
\eqa{
\hP_i^{S=3/2}=\frac{2}{3}\( \hbS_{i-1}\cdot \hbS_i+\hbS_i\cdot \hbS_{i+1}+\hbS_{i-1}\cdot \hbS_{i+1}\) +\frac{1}{2}.
}{P3/2}

\figin{8.5cm}{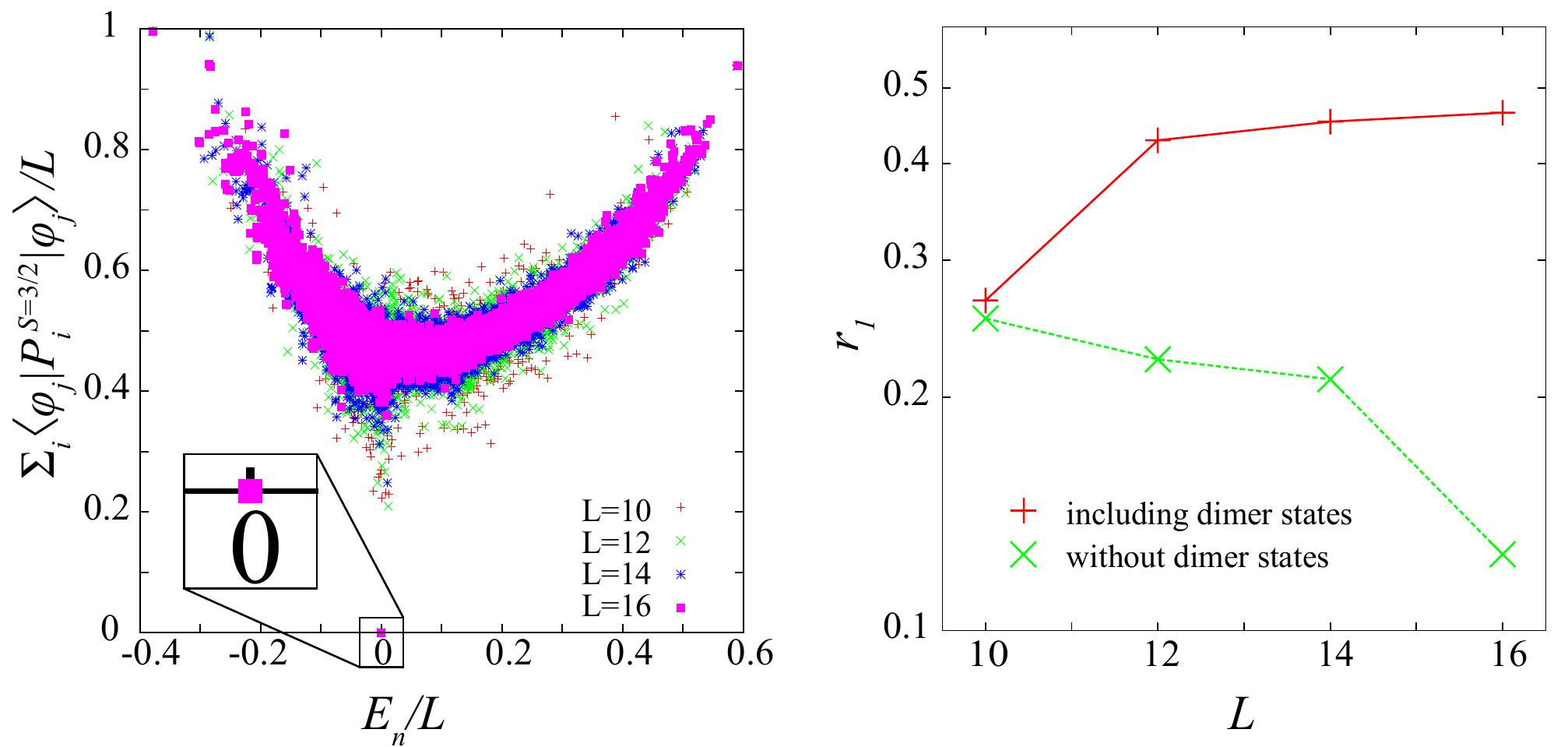}{
(Color online)
The model 1 with the Hamiltonian given by \eref{H1} is investigated.
Left: The plot of expectation values of $(1/L)\sum_i \hP_i^{S=3/2}$ for all energy eigenstates.
The color represents the system size.
The outlier point $(0,0)$ corresponds to the two embedded dimer states.
Right: The most outlying value of $(1/L)\sum_i \hP_i^{S=3/2}$ from its microcanonical average denoted by $r_1$ for all energy eigenstates (red) and those except two dimer states (green).
}{model1}

The analyses on the Majumdar-Ghosh model~\cite{MG}, whose Hamiltonian is $\hH_{\rm MG}:=\sum_i \hP_i^{S=3/2}$,  tell that $\hH_{\rm MG}$ has two dimer states as its ground states with zero energy:
\eq{
\ket{\Psi_{\rm MG}^1}:=\prod_{n=1}^{L/2}\ket{v_{2n-1,2n}}, \ \ket{\Psi_{\rm MG}^2}:=\prod_{n=0}^{L/2-1}\ket{v_{2n,2n+1}}, 
}
where $\ket{v_{i,j}}$ is the valence-bond (spin singlet):
\eq{
\ket{v_{i,j}}:=\frac{1}{\sqrt{2}}(\ket{\up}_i\ket{\down}_j-\ket{\down}_i\ket{\up}_j).
}
Because the total angular momentum of a spin-singlet is zero, the total angular momentum of three spins including a spin-singlet pair is always 1/2, which implies that these two states are ground states of $\hH_{\rm MG}$: $\hH_{\rm MG} \ket{\Psi_{\rm MG}^i}=0$ ($i=1,2$).
In addition, it is known that the ground states are only these two~\cite{CEM, AKLT}.

By setting $\hH'=0$, $\{\hh_i\}$ as translation invariant local terms, and tuning the origin of $\hh_i$ properly, the Hamiltonian
\eqa{
\hH_1:=\sum_{i=1}^L \hP_i^{S=3/2} \hh_i \hP_i^{S=3/2}
}{H1}
has two dimer states $\ket{\Psi_{\rm MG}^i}$ ($i=1,2$) as its energy eigenstates with zero energy, which settles in the middle of the energy spectrum.
These two dimer states span the Hilbert subspace $\mathcal{T}$, and they do not represent a thermal state of this Hamiltonian~\cite{SM}.
Hence, this model is a counterexample to the ETH.
It is worth noting that this model in general satisfies the conditions (i)-(iii)~\cite{LCQ}.
In particular, we emphasize that a local projection operator $\hP_i^{S=3/2}$ is \textit{not} a local conserved quantity.

Numerical calculations reveal that the two dimer states are not thermal, but all the other eigenstates are thermal.
We set the local Hamiltonian $\hh_i$ as
\balign{
\hh_i:=\sum_{\alpha=x,y,z}[ &J_\alpha (\Sa_{i-1}\Sa_i+\Sa_i\Sa_{i+1}) \nt \\
&+J'_\alpha (\Sa_{i-2}\Sa_i+\Sa_i\Sa_{i+2})-h_\alpha \Sa_i] \lb{hi1}
}
with $J_x=J_y=1$, $J_z=-0.6$, $J'_x=-0.8$, $J'_y=J'_z=0$, $h_x=0.3$, $h_y=0$, $h_z=0.1$.
Full diagonalization results of $\sum_i \bra{\phi_j}\hP_i^{S=3/2}\ket{\phi_j}/L$ versus energy density $E_j/L$ for all energy eigenstates are depicted in the left panel of \fref{model1}.
The outlying point (0,0) corresponds to the two degenerate dimer states.
We see that except these two states the fluctuation reduces as increasing system size, which is consistent with the claim that other energy eigenstates are thermal.

The ETH is numerically studied by considering the following indicator:
\eqa{
r[\hat{O}] :=\max_{j}\abs{\bra{\phi_j}\hat{O}\ket{\phi_j}-\langle \hat{O}\rangle _{\rm mc}^{E_j, \Di E}},
}{r}
where $\hat{O}$ is a local observable and $j$ runs all possible energy eigenstates in some fixed range of the energy density.
$\la \cdot \ra _{\rm mc}^{E_j, \Di E}$ represents the ensemble average in the microcanonical ensemble with energy between $E_j-\Di E$ and $E_j$.
If $r$ tends to zero as the system size increases, it implies that the ETH is satisfied.

Here, for the Hamiltonian $\hat{H}_1$ and $\hat{O}=(1/L)\sum_{i=1}^L\hP_i^{S=3/2}$~\cite{degeneracy}, we compute $r$ in the energy range $-0.1\leq E_j/L\leq 0.1$, which we call $r_1$.
The microcanonical energy width is set as $\Delta E=0.01\sqrt{L}$.
In the right panel of \fref{model1}, we plot $r_1$ versus system size $L$ for all eigenstates (red) and all eigenstates except two dimer states (green).
The former does not decrease with increase of $L$, while the latter indeed does, which is expected behavior for thermal eigenstates.
Our numerical results clearly show that the Hamiltonian $\hH_1$ has two nonthermal eigenstates $\ket{\Psi_{\rm MG}^i}$ ($i=1,2$) and $2^L-2$ thermal eigenstates.

\para{Model 2: exponentially-many nonthermal states}
We can also embed exponentially-many target states (i.e., $N_{\rm ex}=O(e^L)$).
We demonstrate this through constructing the second counterexample to the ETH.
Consider a one-dimensional spin chain of $S=1$ with length $L$ with the periodic boundary condition.
The state of each spin is given by a linear combination of three eigenstates of $\hS^z$ expressed as $\ket{1}$, $\ket{0}$ and $\ket{-1}$.
The label of sites is same as that in the model 1.
We now introduce a projection operator to the subspace with $S_i^z=0$ as $\hP_i^0:=1-(\hS_i^z)^2$ and its compliment as $\hQ_i:=1-\hP_i^0=(\hS_i^z)^2$.
Using this, we define a non-local operator $\hcQ:=\prod_{i=1}^L \hQ_i$, which takes 1 if and only if all spins are linear combinations of $\ket{\pm 1}$.
We also introduce pseudo Pauli matrices between two states $\ket{1}$ and $\ket{-1}$ defined as
\balign{
\ts^x&:=\ket{1}\bra{-1}+\ket{-1}\bra{1}, \\
\ts^y&:=-i\ket{1}\bra{-1}+i\ket{-1}\bra{1}, \\
\ts^z&:=\ket{1}\bra{1}-\ket{-1}\bra{-1}.
}

With noting that $[\hP_i^0, \hh_{i-1,i+1}]=0$ and $[\hP_i^0, \hH']=0$ are satisfied for $\hh_{i-1,i+1}$ with its support $\{ i-1, i+1\}$ and $\hH'$ as a function of $\ts^\alpha$ ($\alpha=x,y,z$), we construct a Hamiltonian
\eqa{
\hH_2:=\sum_{i=1}^L \hh_{i-1,i+1}\hP_i^0+\hH' ,
}{H2}
where we used a relation $\hP_i\hh_i\hP_i=\hh_i\hP_i$ for $[\hh_i, \hP_i]=0$.
This Hamiltonian has $2^L$ eigenstates in the subspace with $\calQ=1$, and these eigenstates span the Hilbert subspace $\mathcal{T}$.
With the same discussion for the model 1, we conclude that the model 2 also violates the ETH even though it satisfies the conditions (i)-(iii)~\cite{LCQ}.

\figin{8.5cm}{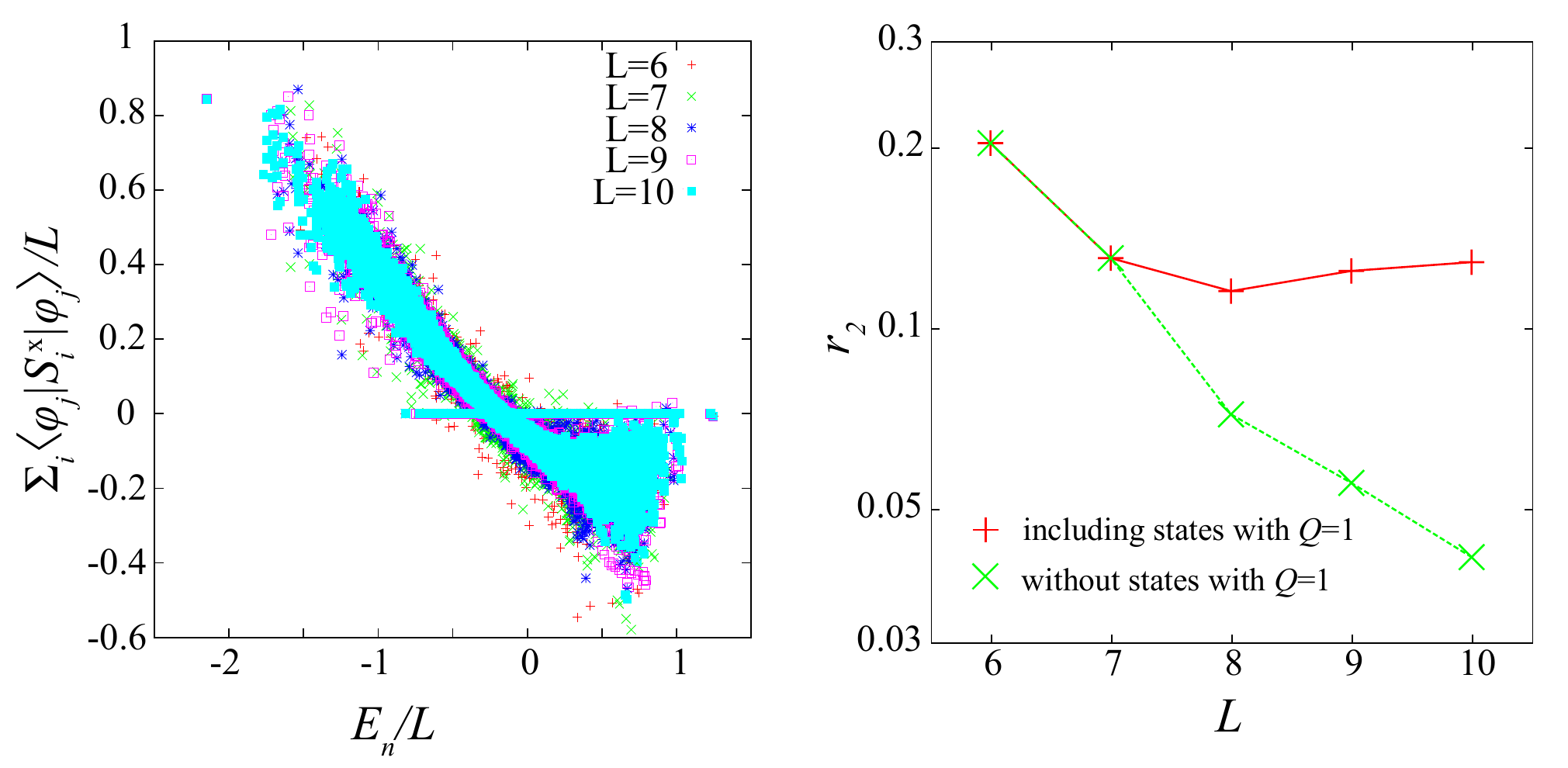}{
(Color online)
The model 2 with the Hamiltonian given by \eref{H2} is investigated.
Left: The plot of expectation values of $(1/L)\sum_i \hat{S}_i^x$ for all energy eigenstates.
The color represents the system size.
At $\sum_i \bra{\phi_j}\hS_i^x\ket{\phi_j}/L=0$, we see a horizontal bar, which corresponds to the embedded states with $\calQ=1$.
Right: The most outlying value of $(1/L)\sum_i \hS_i^x$ from its microcanonical average denoted by $r_2$ for all energy eigenstates (red) and those except states with $\calQ=1$ (green).
}{model2}

Numerical calculations reveal that the eigenstates with $\mathcal{Q}=1$ are not thermal, while the eigenstates with $\mathcal{Q}=0$ are thermal.
We set the local Hamiltonian $\hh_{i-1,i+1}$ and $\hH'$ as
\balign{
\hh_{i-1,i+1}=&\sum_{\alpha=x,y,z}J_\alpha \hS_{i-1}^\alpha \hS_{i+1}^\alpha -h_\alpha (\hS_{i-1}^\alpha +\hS_{i+1}^\alpha)+D \lb{hh2} \\
\hH'=&\sum_{i=1}^L \sum_{\alpha=x,y,z} J'_\alpha \ts^\alpha_i \ts^\alpha_{i+1}-h'_\alpha \ts^\alpha_i \lb{H'}
}
with $J_x=-0.8$, $J_y=0.2$, $J_z=0.4$, $h_x=1$, $h_y=0$, $h_z=0.3$, $D=-0.4$, $J'_x=-0.6$, $J'_y=0.4$, $J'_z=0.8$, $h'_x=h'_y=0$, $h'_z=-0.2$.
First, we compute the expectation value of $x$-component of the spin per site $ \sum_i \bra{\phi_j}\hS_i^x\ket{\phi_j}/L$ versus energy density $E_j/L$, which is plotted in the left panel of \fref{model2}.
The horizontal bar at $\sum_i \bra{\phi_j}\hS_i^x\ket{\phi_j}/L=0$ corresponds to the embedded eigenstates with $\calQ=1$.
Except these embedded states, the fluctuation reduces as increasing system size, which is consistent with the claim that all energy eigenstates with $\calQ=0$ are thermal.

We also consider the indicator of the ETH defined by \eref{r} for the Hamiltonian $\hat{H}_2$ with $\hat{O}=(1/L)\sum_i \hat{S}_i^x$, which we call $r_2$.
Here, $j$ in \eref{r} runs all possible energy eigenstates with $-0.5\leq E_j/L\leq 0$ and $\Di E$ is set to $0.1\sqrt{L}$.
The results are depicted in the right panel of \fref{model2} for all eigenstates (red) and all eigenstates except the embedded eigenstates with $\calQ=1$ (green).
These plots ensure that all eigenstates except the $2^L$ embedded ones are thermal, while the ETH is not satisfied.
Here, we have presented the result for the special choice of $\hat{O}=(1/L)\sum_i \hat{S}_i^x$, but the same conclusion is confirmed for other choices such as $\hat{O}=(1/L)\sum_i \hat{S}_i^z\hat{S}_{i+1}^z$.

We note that although there are exponentially-many nonthermal states $N_{\rm ex}=O(e^L)$, the weak-ETH still holds because the fraction of the nonthermal states is exponentially-small: $N_{\rm ex}/N_{\rm tot}=(2/3)^L=O(e^{-L})$.
The weak-ETH says that the variance of a local observable $\hO$ defined as
\eq{
V(\hO ):=\frac{1}{N_{[-0.5,0]}}\sum_j \( \bra{\phi_j}\hO\ket{\phi_j}-\langle \hO \rangle _{\rm mc}^{E_j, \Di E}\) ^2
}
converges to zero in the thermodynamic limit $L\to \infty$, where $j$ runs all possible energy eigenstates with $-0.5\leq E_j/L\leq 0$ and the number of such energy eigenstates is denoted by $N_{[-0.5,0]}$.
Our model violating the ETH shows the exponential decay of the standard deviation $V^{1/2}$ of $(1/L)\sum_i \hat{S}_i^x$ and $(1/L)\sum_i \hat{S}_i^z\hat{S}_{i+1}^z$ (see \fref{model2-2}).
This means that the exponential decay of $V(\hO)$ with respect to $L$ does not necessarily imply the ETH, which is contrary to the previous argument~\cite{BMH}.

\figin{4.5cm}{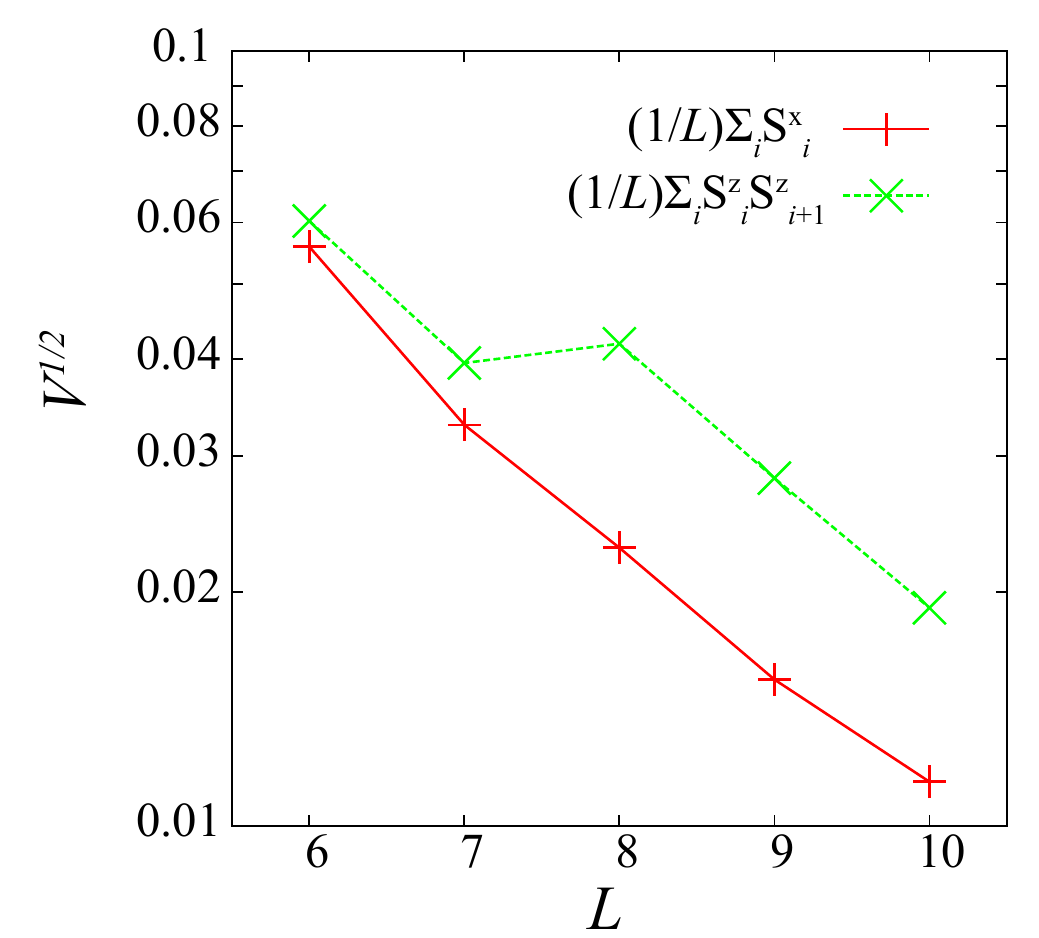}{
(Color online)
The standard deviation $V^{1/2}$ of $(1/L)\sum_i \hat{S}_i^x$ (red) and $(1/L)\sum_i \hat{S}_i^z\hat{S}_{i+1}^z$ (green) in the model 2 with respect to the system size $L$.
Both of them show exponential decay.
}{model2-2}

\para{Thermalization}
All existing models without the ETH including integrable systems and many-body localization generally do not thermalize after a quench.
This is why some researchers believe that the ETH is essential for thermalization.
However, we shall show a good reason to consider that our models indeed thermalize after a physically-plausible quench, which we refer to as a quench from a system with finite temperature.

We take the model 2 as an example.
If all the sites $i$ are in the sector of $Q_i=1$ in the initial state, the system does not thermalize.
However, since all the eigenstates with $\mathcal{Q}=0$ are thermal, we claim that even a single defect of $Q_i=0$ is enough to thermalize the system.

Consider a quench to $\hH_2$ from a thermal state of another Hamiltonian denoted by $\ket{\psi_{\rm ini}}$.
As shown in Supplemental Material, the expectation value of $(1/L)\sum_{i=1}^L\hP^0_i$ in a thermal state is strictly positive, and its variance converges to zero in the thermodynamic limit.
In contrast, all the embedded eigenstates always take zero with the measurement of $(1/L)\sum_{i=1}^L\hP^0_i$, which implies that $\ket{\psi_{\rm ini}}$ has quite a small weight on the nonthermal embedded states, and the system must thermalize.
We, however, note that for relatively small system size, the embedded states with $\calQ=1$ can have relatively large weight, and in that case the system does not thermalize.

\para{Discussion}
We proposed a systematic procedure to construct models with the conditions (i) translation invariance, (ii) no local conserved quantity, and (iii) local-interaction, but not satisfying the ETH, contrary to the common belief.
Our method enables us to embed the target states as energy eigenstates of the Hamiltonian with (i)-(iii) in the middle of energy spectrum, and these embedded states are generally nonthermal~\cite{SM}.
One advantage of our approach lies in the fact that the violation of the ETH is analytically proven, in contrast to numerical simulations which are inevitably affected by the finite size effect (see also a series of discussion on the Ising model with both longitudinal and transverse magnetic fields~\cite{KIH, BCH, Kim}, where slowly decaying observables disturb accesses with numerical simulations).

On the basis of the numerical result that all the energy eigenstates except embedded states are thermal, we argued that the constructed models will thermalize after a physically-plausible quench.
The presence of nonthermal energy eigenstates implies the existence of an initial state which fails to thermalize~\cite{Pal}, but we do not expect to pick up such an initial state through a finite-temperature quench for sufficiently large system sizes since even a single defect can completely recover the thermal property.
Our results elucidate the fact that the role of thermal energy eigenstates in the mechanism of thermalization is not so simple than expected.

The second model contains exponentially-many nonthermal states, which is usually expected to be a property of integrable systems.
Our result implies that the number of nonthermal states does not determine the fate of the presence/absence of thermalization.
To understand thermalization, the property of preparable initial states (e.g., weight to nonthermal eigenstates) should also be taken into consideration.

Apart from the study on thermalization, our procedure sounds a fruitful methodology to obtain interesting Hamiltonians.
Our procedure can embed any state which is a ground state of a frustration-free Hamiltonian.
Both the matrix-product states (MPS) and the projected-entangled-pair states (PEPS) are known to be written as a ground state of a frustration-free Hamiltonian~\cite{FNW, SCG}, and thus they can be embedded to the middle of the energy spectrum of a many-body Hamiltonian.
Our method opens the way to import many brilliant achievements on the ground state of quantum systems to thermal (excited) states.

\para{Acknowledgement}
We wish to thank Takahiro Sagawa, Hal Tasaki, Eiki Iyoda, Sho Sugiura and Masaki Oshikawa for fruitful and stimulating discussion and many constructive comments.
We also thank Akira Shimizu and Ryusuke Hamazaki for helpful advice.
We thank Terry Farrelly for informing us of the reference~\cite{Pal}.
NS is supported by Grant-in-Aid for JSPS Fellows JP17J00393, and TM is supported by  JSPS KAKENHI Grant No. 15K17718.

\clearpage

\makeatletter
\long\def\@makecaption#1#2{{
\advance\leftskip1cm
\advance\rightskip1cm
\vskip\abovecaptionskip
\sbox\@tempboxa{#1: #2}%
\ifdim \wd\@tempboxa >\hsize
 #1: #2\par
\else
\global \@minipagefalse
\hb@xt@\hsize{\hfil\box\@tempboxa\hfil}%
\fi
\vskip\belowcaptionskip}}
\makeatother
\newcommand{\vo}{\upsilon}
\newcommand{\midskip}{\vspace{3pt}}

\setcounter{equation}{0}
\def\theequation{A.\arabic{equation}}

\begin{widetext}

\begin{center}
{\bf \Large Supplemental Materials for ``Systematic Construction of Counterexamples to Eigenstate Thermalization Hypothesis"}

\bigskip
Naoto Shiraishi and Takashi Mori
\end{center}

\bigskip

\begin{quotation}
Here we shall describe in detail the technical part of the main text, which includes definitions of some important notions and the proof of the violation of the eigenstate thermalization hypothesis in the class of models considered in the main text.
We will use the same notation as that in the main text.
\end{quotation}

\bigskip

\bigskip\noindent
{\bf \large Definition of the thermalization}
\midskip

We here give the precise definition of {\it thermalization}.
We first define {\it  thermalization with respect to an observable $\hO$}.
We here restrict the class of observables $\hO$ to local observables and density of macroscopic observables.

We consider a sequence of Hamiltonians $\{ \hH^L\}_L$, where $L$ is the system size and $L$ goes to infinity.
Let $\rho_0^L$ be an initial density matrix for a system with size $L$, and $\rho^L(t)=e^{-\frac{i}{\hbar}\hH^L t}\rho_0^L e^{\frac{i}{\hbar}\hH^L t}$ be a state at time $t$ under the time evolution with the Hamiltonian $\hH^L$.
Then, we say that the system with the initial state $\rho^L_0$ thermalizes with respect to an observable $\hO$ if 
\eqa{
\lim_{L\to \infty} \lim_{T\to \infty} \frac{1}{T}\int_0^T dt \chi \( \abs{\Tr[\hO \rho^L(t)]-\langle \hat{O}\rangle _{\rm mc}^{L, \Tr[\hH^L\rho_0^L], \Di E} }>\ep \) =0
}{def-thermali}
holds for any positive $\ep>0$.
Here, $\chi (\cdot )$ takes one if the statement inside the clause is true and takes zero otherwise, and $\langle \hat{O}\rangle _{\rm mc}^{L, E', \Di E}$ represents the ensemble average in the microcanonical ensemble with the energy shell $E'-\Di E\leq E\leq E'$ with properly-chosen width of energy shell $\Di E$.
The definition of thermalization \eqref{def-thermali} claims that the expectation value of $\hO$ almost always stays near the ensemble average in its corresponding microcanonical ensemble.

If the system with the initial state $\rho^L_0$ thermalizes with any  local observable and density of macroscopic observable $\hO$, we simply say that the system with the initial state $\rho^L_0$ thermalizes.
If one states ``the system thermalizes" without specifying the initial state, its implicit meaning is that the system thermalizes with any physically preparable initial state.

\bigskip

\bigskip\noindent
{\bf \large Definition of the eigenstate thermalization hypothesis}
\midskip

We here give the precise definition of the eigenstate thermalization hypothesis (ETH).
We first define {\it  the ETH with respect to an observable $\hO$}.
We here again restrict the class of observables $\hO$ to local observables and density of macroscopic observables.

We consider a sequence of Hamiltonians $\{ \hH^L\}_L$, where $L$ is the system size and $L$ goes to infinity.
Our two examples of Hamiltonians $\hH_1$ and $\hH_2$ (\eqref{H1} and \eqref{H2} in the main text) naturally arise such sequences by changing the system size $L$.
We denote the $j$-th energy eigenvalue and the corresponding energy eigenstate of $\hH^L$ by $E_j^L$ and $\ket{\phi _j^L}$, respectively.
We then introduce the indicator of $\hO$ with energy density $e_1\leq E/L\leq e_2$ (as \eref{r} in the main text):
\eqa{
r_{e_1,e_2}^L [\hat{O}] :=\max_{j: e_1\leq E_j/L\leq e_2}\abs{\bra{\phi_j^L}\hat{O}\ket{\phi_j^L}-\langle \hat{O}\rangle _{\rm mc}^{L, E_j, \Di E}},
}{r-gen}
which quantifies the divergence of the most outlying value from the corresponding microcanonical ensemble.

If the indicator converges to zero in the thermodynamic limit
\eq{
\lim_{L\to \infty}r_{e_1,e_2}^L [\hat{O}]=0,
}
we say that the Hamiltonian $\hH^L$ (more precisely, the sequence of Hamiltonians $\{ \hH^L\}_L$) satisfies the ETH with respect to an observable $\hO$ with energy density $e_1\leq E/L\leq e_2$.
If  the ETH with energy density $e_1\leq E/L\leq e_2$ is satisfied with all local observables and density of macroscopic observables, we simply say that the Hamiltonian $\hH^L$ satisfies the ETH with energy density $e_1\leq E/L\leq e_2$.
In many papers, the condition for energy density is conventionally dropped, whose implicit meaning is that the ETH is satisfied in the middle of the energy spectrum.

\bigskip

\bigskip\noindent
{\bf \large Rigorous proof of violation of the eigenstate thermalization hypothesis}
\midskip


We consider the Hamiltonian
\eq{
\hH=\sum_i\hP_i\hh_i\hP_i+\hH' ,
}
which is same as \eref{defH} in the main text.
It is assumed that $\hH'$ is a local Hamiltonian which is expressed as a summation of local terms $\{\hh_i'\}$:
\eq{
\hH'=\sum_i\hh_i'.
}

An energy eigenstate $|\phi_j\rangle$ with an eigenenergy $E_j$ is said to be thermal if any local observable $\hO$ satisfies
\eq{
\langle\phi_j|\hO|\phi_j\rangle\simeq\langle \hO \rangle _{\rm mc}^{E_j, \Di E},
}
where $\simeq$ means that both-hand sides converge to the same value in the thermodynamic limit.
What is shown here is that, in general, target states are not thermal, and hence, the eigenstate thermalization hypothesis does not hold.

We set $\hO=\hP_i$ for a fixed $i$.
First, we will show that
\eqa{
\langle\hP_i\rangle_{\rm mc}^{E,\Delta E}\simeq\langle\hP_i\rangle_{\rm can}^{\beta(E)}
\geq c_{\beta(E)}
}{bound}
with a temperature-dependent constant $c_{\beta}$, which is strictly positive for any $|\beta|<+\infty$ and independent of the system size.
Here, the expectation value in the canonical ensemble at the inverse temperature $\beta$ is denoted by $\langle\cdot\rangle_{\rm can}^{\beta}$, and $\beta(E)$ is the inverse temperature corresponding to the energy $E$, which is, more precisely, defined by $\langle\hH\rangle_{\rm mc}^{E,\Delta E}=\langle\hH\rangle_{\rm can}^{\beta(E)}$.
The relation $\langle\hP_i\rangle_{\rm mc}^{E,\Delta E}\simeq\langle\hP_i\rangle_{\rm can}^{\beta(E)}$ is a result of the ensemble equivalence between the microcanonical and the canonical ensembles~\cite{MAMW}.
The inequality $\langle\hP_i\rangle_{\rm can}^{\beta}\geq c_{\beta}$ is proved later.

Next, we consider a fixed target state $|\phi_j\rangle\in\mathcal{T}$ with the eigenenergy $E_j$.
We find that $\beta(E_j)$ is in general finite.
Even if $\beta(E_j)=\pm\infty$, by considering a modified Hamiltonian with $\hh_i\rightarrow\hh_i+\lambda\hat{1}$ for all $i$, where $\hat{1}$ is the identity operator and $\lambda$ is an arbitrary real number, we can make $|\beta(E_j)|<+\infty$ for some $\lambda$ (in particular, there exists $\lambda$ such that $\beta(E_j)=0$).

By combining these results, it is concluded that, for a target state $|\phi_j\rangle\in\mathcal{T}$, $\langle\phi_j|\hP_i|\phi_j\rangle=0$ but $\langle\hP_i\rangle_{\rm mc}^{E_j,\Delta E}\gtrsim c_{\beta(E_j)}>0$, which clearly shows that the target state is nonthermal.

\bigskip\noindent
{\bf \large Proof of $\langle\hP_i\rangle_{\rm can}^{\beta}\geq c_{\beta}$}

For a fixed $i$, we decompose the Hamiltonian $H$ as
\eqa{
\hH=\hat{H}_i+\hat{V}_i,
}{decomp}
where the support of $\hat{V}_i$ does not overlap with that of $\hP_i$.
Because $\hH$ is written as a summation of local terms with finite operator norm, and because $\hP_i$ is a local operator with a finite support, we can choose $\hat{H}_i$ so that its operator norm $\|\hH_i\|$ is bounded and independent of the system size.

We write $\langle\hP_i\rangle_{\rm can}^{\beta}$ as
\eq{
\langle\hP_i\rangle_{\rm can}^{\beta}=\frac{\mathrm{Tr}\,\hP_ie^{-\beta \hH}}{\mathrm{Tr}\,e^{-\beta \hH}},
}
and bounds of the numerator and the denominator are evaluated separately.
For simplicity, it is assumed that $\beta\geq 0$, but we will finally remark that the similar bound is also obtained for $\beta<0$.

First, we consider the numerator,
\eq{
\mathrm{Tr}\,\hP_i e^{-\beta \hH}=\sum_{n:\langle n|\hP_i|n\rangle=1}\langle n|e^{-\beta \hH}|n\rangle,
}
where $\{|n\rangle\}$ is an arbitrary complete orthonormal basis that diagonalizes $\hP_i$.
The summation is taken over $|n\rangle$ whose eigenvalue of $\hP_i$ is 1.
Using the Peierls inequality~\cite{Pei,Husimi}, $\langle \psi|e^A|\psi\rangle\geq e^{\langle \psi|A|\psi\rangle}$ for any normalized $|\psi\rangle$ and $A$ Hermitian operator, we obtain
\eq{
\sum_{n:\langle n|\hP_i|n\rangle=1}\langle n|e^{-\beta \hH}|n\rangle\geq \sum_{n:\langle n|\hP_i|n\rangle=1}e^{-\beta\langle n|\hH_i|n\rangle-\beta\langle n|\hat{V}_i|n\rangle}.
}
A simple relation $\langle n|\hH_i|n\rangle\leq\|\hH_i\|$ yields the lower bound of the right-hand side as
\eq{
\sum_{n:\langle n|\hP_i|n\rangle=1}e^{-\beta\langle n|\hH_i|n\rangle-\beta\langle n|\hat{V}_i|n\rangle} \geq e^{-\beta\|\hH_i\|}\sum_{n:\langle n|\hP_i|n\rangle=1}e^{-\beta\langle n|\hat{V}_i|n\rangle}.
}
Here we choose $\{|n\rangle\}$ as eigenstates of $\hat{V}_i$, which is possible because $[\hat{V}_i,\hP_i]=0$ and $\{|n\rangle\}$ is an arbitrary complete orthonormal basis diagonalizing $\hP_i$.
Under the above choice of $\{ \ket{n}\}$, 
\eqa{
\sum_{n:\langle n|\hP_i|n\rangle=1}e^{-\beta\langle n|\hat{V}_i|n\rangle}
=\mathrm{Tr}\,\hP_ie^{-\beta\hat{V}_i}
}{num}
is satisfied, and using these relations we obtain
\eqa{
\mathrm{Tr}\,\hP_i e^{-\beta\hH}\geq e^{-\beta\|\hH_i\|}\mathrm{Tr}\,\hP_ie^{-\beta\hat{V}_i}.
}{den}

Next, we consider $\mathrm{Tr}\,e^{-\beta\hH}$.
Using the Golden-Thompson inequality~\cite{Golden,Thom}, $\mathrm{Tr}\,e^{A+B}\leq \mathrm{Tr}\,e^Ae^B$ for Hermitian operators $A$ and $B$, we obtain
\eqa{
\mathrm{Tr}\,e^{-\beta\hH}=\mathrm{Tr}\,e^{-\beta\hH_i-\beta\hat{V}_i}
\leq\mathrm{Tr}\,e^{-\beta\hH_i}e^{-\beta\hat{V}_i}
\leq e^{\beta\|\hH_i\|}\mathrm{Tr}\,e^{-\beta\hat{V}_i}.
}{GT}

By combining \eref{den} and \eref{GT}, we arrive at
\eq{
\langle\hP_i\rangle_{\rm can}^{\beta}\geq e^{-2\beta\|\hH_i\|}\frac{\mathrm{Tr}\,\hP_ie^{-\beta\hat{V}_i}}{\mathrm{Tr}\,e^{-\beta\hat{V}_i}}.
}
Since the support of $\hP_i$ does not overlap with that of $\hat{V}_i$, it is found that $\mathrm{Tr}\,\hP_ie^{-\beta\hat{V}_i}/\mathrm{Tr}\,e^{-\beta\hat{V}_i}$ is independent of $\beta$ and thus equals to the infinite-temperature average, $\langle\hP_i\rangle_{\rm can}^{\beta=0}$.
In this way, we obtain the lower bound,
\eqa{
\langle\hP_i\rangle_{\rm can}^{\beta}\geq e^{-2\beta\|\hH_i\|}\langle\hP_i\rangle_{\rm can}^{\beta=0}=: c_{\beta}.
}{P_bound}
Since both $\|\hH_i\|$ and $\langle\hP_i\rangle_{\rm can}^{\beta=0}$ are strictly positive and independent of the system size, $c_{\beta}$ is also strictly positive and independent of the system size as long as $\beta<+\infty$.

Finally, we remark that for $\beta<0$, the same bound as \eref{P_bound} with $|\beta|$ instead of $\beta$ is obtained, and hence $\langle\hP_i\rangle_{\rm can}^{\beta}$ is strictly positive for any $\beta$ with $|\beta|<+\infty$.

\clearpage

\end{widetext}

\end{document}